# Ultra-low Power Microwave Oscillators based on Phase Change Oxides as Solid-State Neurons

B. Zhao,[1] J. Ravichandran,[1,2*]

[1]*Mork Family Department of Chemical Engineering and Materials Science, University of Southern California, Los Angeles, California, 90089, USA*
[2]*Ming Hsieh Department of Electrical Engineering, University of Southern California, Los Angeles, California, 90089, USA*

Neuro-inspired computing architectures are one of the leading candidates to solve complex, large-scale associative learning problems. The two key building blocks for neuromorphic computing are the synapse and the neuron, which form the distributed computing and memory units. Solid state implementations of these units remain an active area of research. Specifically, voltage or current controlled oscillators are considered a minimal representation of neurons for hardware implementations. Such oscillators should demonstrate synchronization and coupling dynamics for demonstrating collective learning behavior, besides the desirable individual characteristics such as scaling, power, and performance. To this end, we propose the use of nanoscale, epitaxial heterostructures of phase change oxides and oxides with metallic conductivity as a fundamental unit of an ultra-low power, tunable electrical oscillator capable of operating in the microwave regime. Our simulations show that optimized heterostructure design with low thermal boundary resistance can result in operation frequency of up to 3 GHz and power consumption as low as 15 fJ/cycle with rich coupling dynamics between the oscillators.

## I. INTRODUCTION

As we approach the scaling limits of the digital logic based on von Neumann architecture [1], the available large computational power can be leveraged to solve associative learning problems such as pattern matching, image recognition [2, 3]. Alternative paradigms such as neuromorphic computing is considered efficient approaches to solve these problems and can lead to the proliferation of distributed and autonomous sensing, computing and communication platforms [4, 5]. Neuromorphic computing incorporates computation based on biological brain-inspired mechanisms, where the time series of signals are used to both modulate the memory elements and perform computations. This approach uses an interconnected network of neurons and synapses, where the neurons perform the distributed analog computing and the synapses act as the learning and memory units [6]. Early work on the hardware development based on neuromorphic computing relied heavily on the matured CMOS based platform [7]. However, recently approaches based on spin-torque [8, 9], metal-insulator-transition [10, 11], resistive switching [12], and ferroelectric phase change materials [13, 14] are being explored as alternatives to the traditional CMOS based ring oscillators.

---
[*]jayakanr@usc.edu

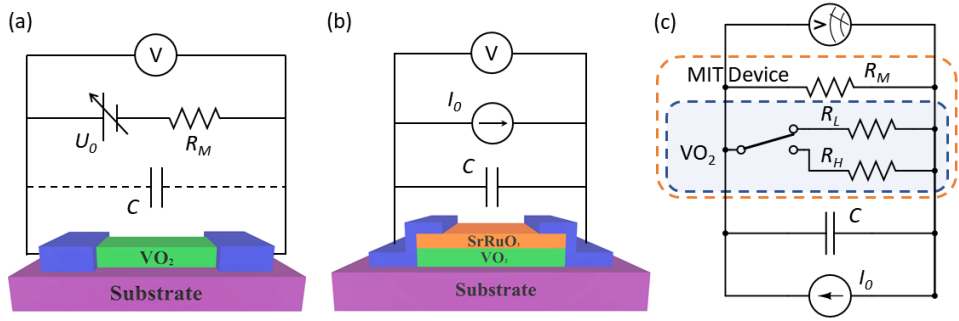

**FIG. 1. Device structure and circuit diagram.** (a) Schematic of the typical VO$_2$ electrical oscillator, where the VO$_2$ channel is connected in series to a known resistor and the Joule heat delivered to the series resistor adds to the extraneous power consumption, (b) Schematic of the proposed VO$_2$ based bilayer electrical oscillator with SrRuO$_3$ as the second layer. The bilayer is connected in parallel to a capacitor and current source is used to inject current into the device. (c) Schematic of the circuit diagram for the proposed oscillator device.

Phase change materials showing metal-insulator transition (MIT), especially VO$_2$[15], have been widely explored as a platform to achieve electrical oscillations [15, 16], often with oscillation frequency up to 9 MHz [17] and operation voltage below 1 V [18]. VO$_2$ undergoes a concomitant first-order structural [19] and electrical transition [20] with a large 5 orders of magnitude change in the electrical conductivity at 68°C in bulk single crystals [21]. Although the exact nature of the phase transition is still under considerable investigation [22-27], this large change in electrical conductivity [28] close to room temperature offers a novel method to create a feedback loop for switching between the metallic and insulating states, and hence, create electrical oscillations. The prototypical implementation of such an oscillator involves a VO$_2$ channel with lateral contacts connected to a series resistor and an external bias [29] (Fig. 1 (a)). In the insulating state, the Joule heating of the VO$_2$ channel effects a transformation to the metallic state. In the metallic state, the Joule heating is regulated by the external resistor and the available heating power is lower due to the low resistivity of the metallic state. This leads to cooling of the channel and recovery of the insulating state in VO$_2$. The highest oscillation frequency achieved in such devices was as high as 1 MHz with a power consumption of 1 nJ/cycle [30]. Recently, out-of-plane contact geometry to VO$_2$ channel was explored with a successful increase in the oscillator frequency to 9 MHz with a lower power consumption of 0.5 nJ/cycle [17]. A variation from both these geometries is the use of metal/VO$_2$ bilayer strips as oscillators [31]. Early attempts at using this geometry to induce electrical oscillations explored Pt/VO$_2$ bilayers [32]. These devices suffered from failure during repeated cycling and large thermal resistance at the Pt/VO$_2$ interface.

Several theoretical and experimental studies have shown that the time constant necessary for the metal-insulator transition can be as low as 2 ns and hence, one can achieve oscillation frequencies much higher than the observed 9 MHz in VO$_2$ based electrical oscillators [33-35]. Further, the current demonstrations suffer from large power consumption than what is necessary to induce the phase transition, which will limit their utility as low power oscillators [35]. To overcome the limitations of the current device designs discussed earlier, we propose an optimal VO$_2$-based oscillator device structure that can achieve both higher power efficiency and higher frequency without sacrificing any tunability compared to conventional electrical oscillators. Here, we report simulations of the heater-on-channel device structure for VO$_2$ based electrical oscillators (as shown in Fig. 1(b)) with minimal interface thermal boundary resistance to achieve oscillation frequency as large as 3 GHz with a small power consumption of as low as 15 fJ/cycle. Our novel epitaxial bilayer device design minimizes the thermal time constant of the system by substantially reducing the active volume of material undergoing the metal-to-insulator transition. This not only leads to microwave oscillations but also reduces the power consumption drastically compared to earlier designs.

## II. DEVICE STRUCTURE AND MECHANISM OF OPERATION

Our studies employ epitaxial bilayers of $SrRuO_3$ and $VO_2$ etched into a channel with lateral dimensions spanning 0.1-10 μm for length and width, and 5-30 nm for the thickness of each layer. As noted in Fig. 1(b), we will pass a current $I_0$ through the channel with a parallel capacitor with capacitance $C$ to set up the oscillator. We carried out phenomenological modeling of the device operation to establish its performance metrics such as oscillation frequency, power consumption. The device is represented as a parallel bilayer system mounted on a semi-infinite substrate as the bilayer thickness is expected to be at least 3-4 orders of magnitude smaller than the substrate thickness. The phenomenological model consists of two constitutive equations that address the conservation of charge and heat energy in the system [31]. The conservation of charge is represented as a first-order differential equation as noted below:

$$I_0 = CdV/dt + V/R_{MIT} + V/R_M \qquad (1)$$

where $R_M$ and $R_{MIT}$ are the resistances of the metallic oxide and $VO_2$ layer, and $V$ is the voltage across the device. The equation representing the Newton's law of cooling (or the conservation of heat energy) is given below:

$$mC_m dT/dt = V^2/R_{MIT} + V^2/R_M - \kappa(T - T_0) \qquad (2)$$

where $m$, $C_m$ are the mass and heat capacity of the MIT device, $T$, $T_0$ are the temperature of the device and the ambient respectively, $\kappa$ is the heat conductance between the device and the substrate. To solve these equations in a self-consistent manner, we have created a one-to-one map of the temperature and the resistance state of the $VO_2$ layer. As $VO_2$ undergoes a first-order transition, the specific resistance state of the material is dependent on whether channel is subject to the heating or cooling cycle (for detailed description please see Appendix A). Hence, we have ensured that this temperature-resistance map accounts for the hysteretic nature of the temperature dependence of resistance.

We performed finite element simulations of the temperature profile for the device channel to learn about the heat dissipation pathways and dominant thermal resistances. This analysis was used to identify the relevant contributions to the heat capacity and thermal resistance terms discussed in equation (2) (for detailed description please see Appendix A). Based on these calculations, we concluded that the dominant heat loss from the channel occurs *via* the substrate and the metal contacts. For the optimal device geometry with small heat capacity and small thermal resistance, one can achieve fast heating and cooling dynamics, which are the dominant rate-limiting steps for achieving fast electrical oscillations. We achieve this using a nanoscale mesa of ultrathin bilayers, where heat transfer is not fully limited by the thermal boundary resistance between the bilayer. Unless otherwise specified in the following discussions, we list all the relevant geometric and material parameters for the device in Table I.

TABLE I. Geometric dimensions, electrical and thermal properties of the materials used in the simulation

| Property | Value | Property | Value |
|---|---|---|---|
| Thickness of $VO_2$ | 5 nm | Density of $SrRuO_3$ | 6490 kg/m$^3$ |
| Thickness of $SrRuO_3$ | 4 nm | Density of $VO_2$ | 4570 kg/m$^3$ |
| Channel Length | 100 nm | Specific Heat of $SrRuO_3$ [36] | 650 J/kg K |
| Channel Width | 100 nm | Specific Heat of $VO_2$ [37] | 690 J/kg K |
| Resistivity of $SrRuO_3$ [38] | $10^{-5}$ Ω m | Thermal conductivity of $VO_2$ [37] | 6 W/m K |
| Resistivity of $VO_2$ (298K) | $10^{-2}$ Ω m | Thermal conductivity of Contacts (Pt) | 70 W/m K |
| Resistivity of $VO_2$ (400K) | $4\times10^{-7}$ Ω m | ITC of $SrRuO_3/VO_2$ [39] | $2\times10^8$ W m$^{-2}$ K$^{-1}$ |
| Temperature for IMT | 345 K | ITC of $SrRuO_3$/Substrate [39] | $1\times10^8$ W m$^{-2}$ K$^{-1}$ |
| Temperature for MIT | 335 K | ITC of Channel/Contacts [40] | $1\times10^7$ W m$^{-2}$ K$^{-1}$ |

Note: IMT is metal-insulator transition; MIT is insulator-metal transition; ITC is the Interfacial thermal conductance.

Fig. 2(a) shows the time dependence of the voltage and temperature of the MIT device for operational parameters that achieve a median sustained oscillation frequency of ~3 GHz. The four regimes of operation are represented in the voltage and temperature oscillations by noting the relevant critical points: A, B, C, and D. A and C coincide with the crossover of the channel across the transition temperature, where the charging and discharging of the capacitors are triggered due to large changes to the channel resistance. B and D correspond to turning points in temperature, where the Joule heating and the heat dissipated are equal. As noted in the inset in Fig. 2(a), the power spectrum of the time-dependent voltage curve shows a peak at 3 GHz, which is the median frequency of the oscillations. The time taken by the device for the heating and cooling cycles were dependent on both the geometric and materials properties of the channel, and the operating current and capacitance values. Hence, one can tune the oscillation characteristics by changing these parameters. A closer look at the voltage oscillations shows that the DC component of these oscillations is minimal due to the large mismatch in the resistance between the metallic oxide and the metallic state of $VO_2$. This mismatch reduces the power consumption to an unprecedentedly low value of 15 fJ/pulse.

We compare the performance metrics such as power, frequency, and footprint of the proposed oscillator scheme with other state-of-the-art electrical oscillators in Table II. The widely used CMOS based oscillators (Ring Oscillators and RC Oscillators) have smaller size and improved tunability compared to bulky LC oscillators, but at the expense of increased power consumption. Resistive switching oscillators (RRAM Oscillators) have significantly smaller footprint compared to these schemes but suffer from lower frequency range and higher power consumption. Spin-torque oscillators are a simple and efficient oscillator scheme and produces highly tunable oscillations in a small footprint but consume large amounts of power. Current $VO_2$ technologies achieve low frequencies (~ MHz) with large power consumption but are very competitive in footprint compared to other schemes. Our proposed device can achieve superior metrics compared to all the other schemes considering the oscillator performance metrics.

TABLE II. Comparison with state-of-the-art oscillators

|  | Frequency (GHz) | Power (mW) | Power (fF/pulse) | Footprint ($\mu m^2$) |
|---|---|---|---|---|
| LC oscillator [41] | 5.4 | 0.5 | ~ 92 | $0.18 \times 10^{6\#}$ |
| Ring Oscillator [42] | 3.35 | 1.45 | ~ 430 | $5600^{\#}$ |
| RC Oscillator [43] | $1.35 \times 10^{-3}$ | $0.92 \times 10^{-3}$ | ~ 680 | $5000^{\#}$ |
| RRAM Oscillator [44] | 0.5 | 0.25 | ~ 500 | $330^{\#}$ |
| Spin Torque Oscillator [9] | 0.33 | ~ 0.05 | ~151 | $0.14^{*}$ |
| Out-of-Phase $VO_2$ [17] | $9 \times 10^{-3}$ | ~ 4.5 | ~ $0.5 \times 10^6$ | ~ $1 \times 10^{6*}$ |
| In-Phase $VO_2$ [30] | $1 \times 10^{-3}$ | ~ 1.0 | ~ $1 \times 10^6$ | $40^{*}$ |
| Pt/$VO_2$ Bilayer [32] | $0.3 \times 10^{-3}$ | ~ 0.9 | ~ $3 \times 10^6$ | $1^{*}$ |
| This Work | 3 | 0.045 | 15 | $0.01^{*}$ |

Note: ~ Calculated or estimated from tables and figures; # Chip size; * Channel size;

One can learn about the mechanism of electrical oscillations by monitoring the Voltage-Current characteristics across the bilayer. In Fig. 2(b), we show the Voltage-Current (V-I) relationship for the oscillator channel, when sustained electrical oscillations are achieved. The plot clearly highlights four distinct regimes that constitute a single oscillation cycle. The regimes are: AB: charging, where supplied current is used to charge the capacitor; BC: heating, where current flows into the channel and heats the metallic layer to push the region of $VO_2$ channel right below the metal into the metallic state; CD: discharge, where the metallic $VO_2$ draws additional current from the capacitor by discharging it; and DA: cooling, where the metallic $VO_2$ doesn't have the necessary Joule heating power to sustain the metallic state, which leads to cooling and transition back to insulating state. These stages are achieved by moving clockwise in the V-I plot and continue in a cyclical manner to create sustained electrical oscillations until the external current supply continues to provide the power for the device.

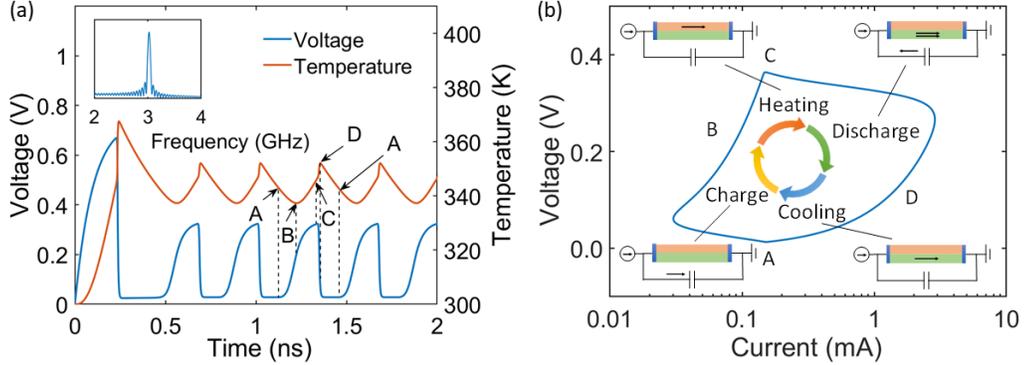

**FIG. 2. Operation mechanism of the bilayer oscillator.** (a) Simulated voltage-current (V-I) curve of the oscillatory cycle during the sustained oscillations in the device; the capacitance used was $C$=1 pF. The four stages of operation of the oscillator are charging, heating, discharging and cooling. The stages are represented in a clockwise manner with the color code used in the V-I curve corresponding to each stage described here. The stages are separated by four critical points: A, B, C and D. A and C represent the onset of the charging and discharging of the capacitor respectively, while B and D represent the onset of heating and cooling; (b) Time dependent variation of voltage and temperature of the MIT device for the highest achievable oscillation frequency of 3 GHz, since the start of the applied current. The different operating critical points within an oscillation are noted as A, B, C, and D as discussed in (a). The inset shows the power spectrum of the voltage oscillations that shows a dominant frequency of 3 GHz.

## III. TUNIBILITY

Next, we discuss the tunability of the oscillator characteristics for the proposed device structure. We have several geometric, materials and operational parameters in our oscillator design that can tune the oscillation characteristics over several orders of magnitude. If one were to mimic the operation of a neuron using these oscillators, it is desirable to achieve tunability of the oscillation frequency, besides low power and high-frequency operation. To elucidate the key operating parameters for this device, we studied the effect of the parallel capacitance and the applied current on the oscillation frequency. The operating regime of the oscillator is represented as a current-capacitance phase diagram in Fig. 3(a). The phase diagram shows four distinct regimes, which are schematically shown. Regions I, II and III show no oscillation, one oscillation, and damped oscillations respectively. Once a threshold in the current and the capacitance is reached, we can enter region IV, where sustained oscillations are possible. In this regime, we can tune the oscillation frequency up to 3 GHz for the device parameters discussed in Table I. It is worth noting that if we modify the device parameters such as the channel area and thicknesses of the two layers in the bilayer, one can achieve finer tunability in the low-frequency regime down to MHz. As the higher frequency of operation in a solid-state neuron can reduce the need for higher scalability, we focus on achieving the highest possible frequency in the proposed device structure. We show the broad tunability of the oscillator frequency with current and the capacitance in Fig. 3(c). One can achieve a large tunability of a factor of 100 for changing capacitance with a constant current and a factor of 4 or more for changing current with constant capacitance. Specific example for these tunability ranges and the corresponding power spectrum are shown in Fig. 7(a) and (b) respectively in the high-frequency regime.

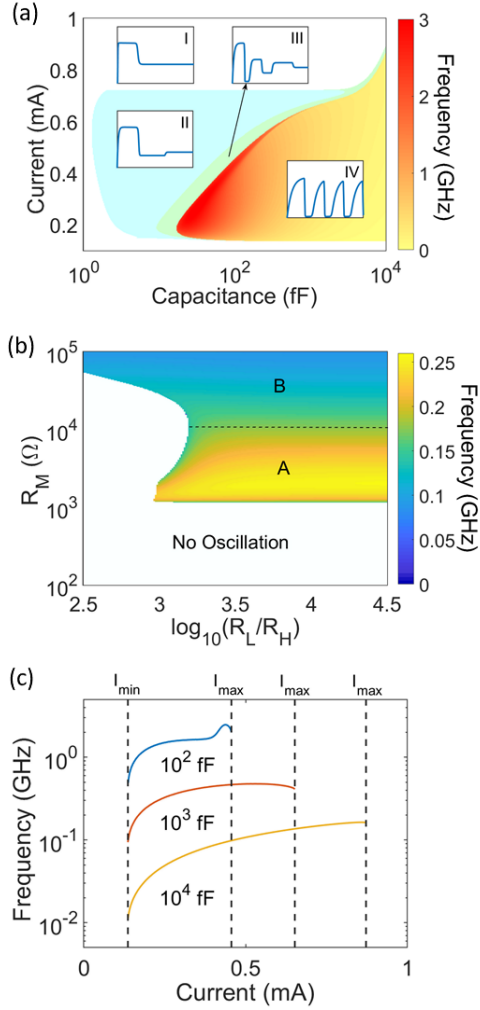

**FIG. 3. Tunability of the bilayer oscillator.** (a) Current-Capacitance phase diagram of the operating regimes of MIT based electrical oscillator. Region IV depicts fully sustained oscillators with the frequency range from 0-3GHz, time evolution of the voltage across the MIT device after the current is applied is characterized by: I – no oscillation; II – one oscillation; III – damped oscillations; IV – sustained oscillations; (b) The effect of the resistances of metallic oxide and MIT layer on the oscillator operation; The dominant parameters controlling the oscillation frequency in region I are $R_M$ and $R_H$, and region II are $R_L$ and $R_H$. ($C$=1000 fF, $I$=0.2mA) (c) Frequency tunability by current and capacitance, $C$=100 fF, 1 pF and 10 pF; current ranges from $I_{min}$ to $I_{max}$.

Another useful parameter to control the oscillation characteristics are the resistances of the individual layers in the bilayer channel. One of the easiest ways to control the resistance is by varying the thickness of the material. Growth conditions can be used to influence the resistivity of the $VO_2$ and $SrRuO_3$ layers too. The amplitude of the voltage oscillations is controlled by the total resistance of the bilayer channel. The voltage across the device for the two cases, where the $VO_2$ is in the insulating state and conducting states are given as:

$$V_{insulating} = I_0 R_M R_L / (R_M + R_L) \qquad (3)$$

$$V_{conducting} = I_0 R_M R_H / (R_M + R_H) \qquad (4)$$

where $I_0$ is the total current, $R_M$, $R_L$, $R_H$ are the resistance of the $SrRuO_3$ layer, the low temperature (300K) insulating state of the $VO_2$ layer, the high temperature (400K) conducting state of the $VO_2$ layer. Fig. 3(b) shows the influence of $R_M$ and the ratio between $R_L$ and $R_H$ on the oscillation frequency. In principle, all the three parameters could be varied, but the influence of the $R_L$ on the oscillation characteristics was minimal, and hence, we chose to fix this value and studied the effect of $R_M$ and $R_H$ on the oscillation frequency. We do not observe any oscillations, when $R_M$ is too small to heat up the device. Similarly, if $R_H$ is higher than or comparable to $R_M$, one cannot establish a feedback loop that sets up the oscillation. The dominant parameters controlling the oscillation process in the two regions I and II are $R_M$ and $R_H$, and $R_L$ and $R_H$ respectively. In region II, we simply recover the device operation for the scheme shown in Fig. 1(a) as $R_M$ is very large and comparable to $R_L$. In other words, the surface layer acts as only a parasitic leakage in region II, but determines the frequency in region I. Clearly, a much higher frequency is achievable in Region I compared to Region II, which emphasizes the advantages of the bilayer device structure over the conventional series resistor-based device structure, provided the heat capacity and the thermal resistances are designed judiciously.

## IV. SYNCHRONIZATION CHARACTERISTICS

Finally, we studied the nature of synchronization between two bilayer electrical oscillators, which is another key characteristic of a solid-state neuron. The coupling between the two electrical oscillators was controlled by a capacitor with capacitance, $C_i$, connecting the two oscillators. For these studies, we used the difference in the current applied to each oscillator as the driving phase difference and the coupling capacitance as the measure of the strength of coupling between the oscillators. We generated the synchronization phase diagram for the oscillators (circuit diagram is shown in Fig. 4(a)), where driving oscillator has the applied current fixed at $I_1$=0.25 mA and the driven oscillator's current $I_2$ varied between 0.15 mA-0.45 mA for varying coupling capacitance ($C_i$=0-100 fF). We mapped the frequency difference between the two oscillators after 10 ns to evaluate their ability to synchronize for a given driving phase and coupling strength and the resultant phase diagram is shown in Fig. 4(b). As noted, large capacitance leads to higher coupling strength and can synchronize oscillators that start with a large driving phase difference. We can also observe interesting synchronization trends for intermediate coupling, which demonstrates the richness of this time dynamics of these coupled oscillators.

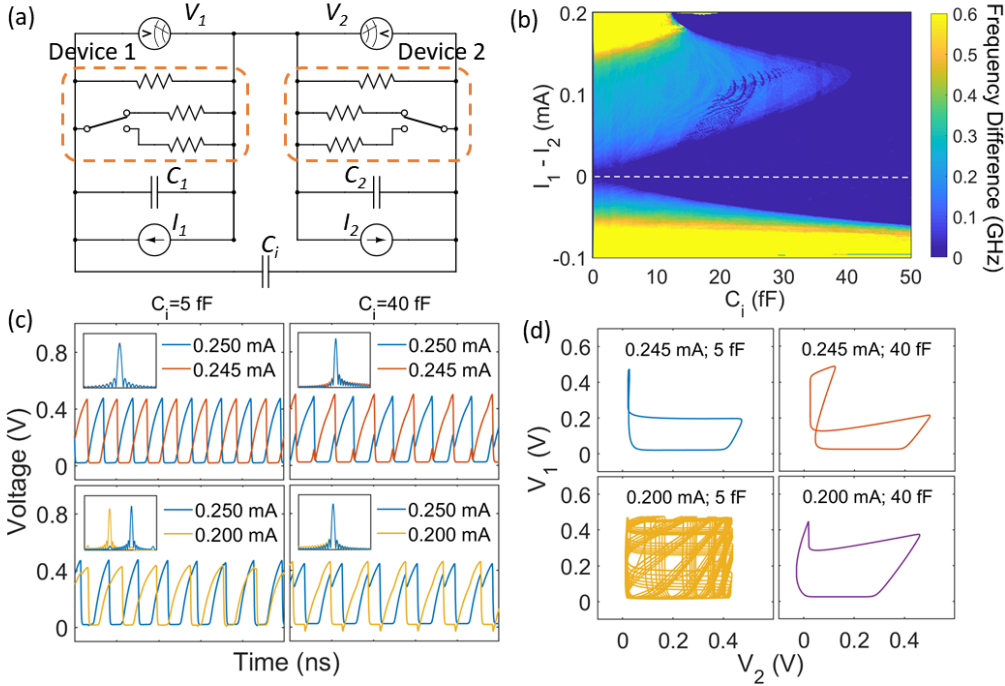

**FIG. 4. Synchronization characteristics of coupled bilayer oscillators.** (a) Schematic of the circuit diagram for two bilayer electrical oscillators connected in series for the synchronization studies; (b) The phase diagram of the synchronization characteristics between the two oscillators for different driving phase (difference in drive current) and coupling strength (coupling capacitance). One can notice the formation of the Arnold's tongue demonstrating the region where synchronization can be easily achieved and the oscillations lock in the same frequency even if they started with different frequencies. (c) Simulated voltage oscillations for two oscillators synchronizing or remaining out of sync with each other. The depicted time span is 10 ns to 15 ns. The inset shows the power spectrum of the oscillations for the two oscillators, which are locked in different frequencies when out of sync. (d) Phase maps for different $I_1$ and $C_i$ combinations corresponding to (c) from 10 ns to 50 ns.

To demonstrate the synchronizing behavior in detail, we show the voltage oscillations for the two oscillators with a different driving phase difference ($I_1$=0.250 mA, $I_2$=0.245 mA; and $I_1$=0.250 mA, $I_2$=0.200 mA) and very different coupling strengths of $C_i$=5 fF and $C_i$=40 fF (Fig. 4(c)). The weak coupling strength of $C_i$=5 fF can synchronize a 0.005 mA current difference between the oscillators but is not sufficient to lock the driven oscillator with 0.050 mA current difference system into a stable oscillatory behavior. Stronger coupling allows one to achieve a more stable synchronized oscillatory behavior for the two oscillators. For example, when $C_i$=40 fF was used to couple the two oscillators, we could achieve synchronization for the driving phase differences of 0.005 mA and 0.050 mA. Phase maps under these conditions are plotted in Fig. 4(d) (time span 10 ns to 50 ns). Clearly unsynchronized phase pattern is shown when $C_i$=5 fF was used to synchronize $I_1$=0.250 mA, $I_2$=0.200 mA. On the other hand, synchronized oscillators show anti-phase relationships in the other three circumstances. A deeper understanding of the nature of this synchronization between coupled oscillators is necessary to design the co-operative learning behavior of the network of oscillators.

## V. CONCLUSION

In summary, we have introduced an ultra-low power, microwave electrical oscillators based on $VO_2$/$SrRuO_3$ bilayer heterostructures. Our proposed device structure is expected to have orders of magnitude higher frequency of operation and lower power consumption than current oscillator device structures. For an optimized device structure, we predict oscillation frequencies as high as 3 GHz and energy assumption as low as 15 fJ/pulse with a large dynamic and static tunability of the oscillator characteristics. The synchronization between two oscillators shows rich and complex dynamical behavior, which is a promising demonstration towards achieving large coupled oscillatory computational systems. This study lays the foundation for future theoretical and experimental studies towards scalable, ultra-lower power solid-state neurons capable of operating in the microwave regime for neuromorphic computing. More broadly, we envision that such ultra-low power tunable microwave oscillators can find application in fields spanning clocking, communication, sensing, and computing.

## ACKNOWLEDGEMENTS

The authors gratefully acknowledge the support from the USC Viterbi School of Engineering startup funds and the Air Force Office of Scientific Research under award number FA9550-16-1-0335. We acknowledge the assistance of Yang Liu and Shanyuan Niu for careful reading of the manuscript and providing useful feedback. We also acknowledge helpful discussions with Asif Islam Khan, Rehan Kapadia, Han Wang, Priya Vashishta and Rajiv Kalia.

## APPENDIX A. METHODS AND SIMULATION APPROACH

### I. Temperature Distribution

We considered the case of an oscillator device etched into a mesa structure and contacted by metal contacts on both the edges (Fig. 1(b) in main text) to learn about the dominant thermal resistances in the system. Due to comparable dimensions of length and width, we assumed that the temperature distribution along the width is uniform. We employed finite element modeling to learn about the temperature distribution along the length of the device (as shown in Fig. 6). We discretized the device into 10 nm (length) x 1 nm (thickness) blocks and each block was assigned a self-consistent temperature, and associated physical properties such as density, specific heat, resistivity, in-plane thermal conductivity and out-of-plane thermal conductivity for each layer (Table I). We included a semi-infinite substrate and 50 nm long contacts with relevant properties. Interfacial thermal conductance was set to the blocks at the interfaces and gave corresponding temperature drops. The temperature of the boundary of the device was fixed to $T_0$=300K. A time domain finite element simulation was carried out in MATLAB, beginning from a uniform temperature $T_0 = 300K$ with a time step $\Delta t = 10\ fs$. Conservation of energy was represented by the following partial differential equation:

$$\rho c_p\ \partial T/\partial t = \nabla \cdot (k \nabla T) + E^2/\rho_r \qquad (5)$$

where $\rho$ is volumetric mass density, $c_p$-specific heat, $k$-thermal conductivity, $E$-electric field, and $\rho_r$ is the resistivity.

### II. Resistance of VO$_2$.

The resistance of VO$_2$ layer ($R_{MIT}$) was obtained by adding the parallel resistances of VO$_2$ under different temperatures according to the temperature profile as shown in Fig. 6. The resistance of each layer was calculated as a function of their temperature as shown in Fig. 5.

### III. Oscillator Characteristics.

The governing equations for the operation of the oscillator are listed as (1) and (2). The initial state is set to be $V = 0, I = 0, T(0) = 300K$. A time domain finite element simulation is carried out with time step $\Delta t = 100\ fs$.

### IV. Synchronization of Coupled Oscillators.

The synchronization between the oscillators was simulated using governing equations that obey Newton's law of cooling and the conservation of charge relationships:

$$I_1 = C_1\ dV_1/dt + V_1/R_{MIT1} + V_1/R_{M1} + C_i\ d(V_1 - V_2)/dt \qquad (7)$$

$$I_2 = C_2\ dV_2/dt + V_2/R_{MIT2} + V_2/R_{M2} + C_i\ d(V_2 - V_1)/dt \qquad (8)$$

Initial states of two oscillators are both set to be $V = 0, I = 0, T(0) = 300K$. Two oscillators are set to have different current input. A time sequence finite element simulation is carried out with time step $\Delta t = 100$ fs.

## V. Resistivity of VO$_2$

Fig. 5 shows the temperature dependence of resistivity of VO$_2$ used in the simulation, which is representative of epitaxial VO$_2$ thin films.[19, 20, 27, 28] Transition temperature during heating was used as 345K (IMT temperature), while that during the cooling was 335K (MIT temperature). Four orders of magnitude change in resistivity were assumed across transition temperature. Resistivity during the heating

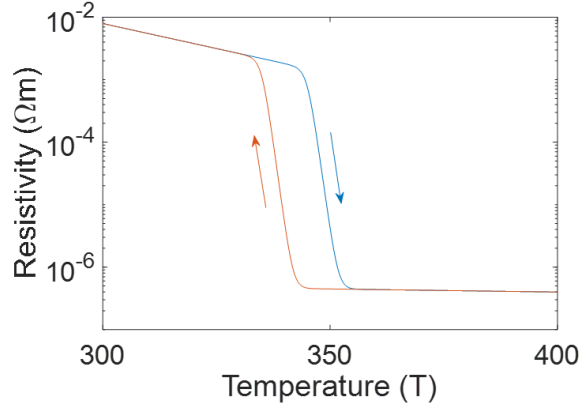

FIG. 5. The temperature dependent resistivity of VO$_2$. We used different transition temperature for heating and cooling cycles as expected for a first order phase transition.

or cooling cycle were considered differently in the simulation.

## VI. Temperature Profile

Fig. 6 shows the simulated temperature map across the cross-section of the MIT channel. As the contact and the substrate are metallic with large conductivity and semi-infinite, their temperature remains almost constant at T=T$_0$ as the channel is heated up. This ensures excellent heat sinking for the heating powers considered in this study and therefore, the temperature profiles and dominant resistances considered remain the same over the whole operating regime. As Fig. 6 indicates, the temperature is uniform across the width and has a parabolic distribution across the length. IMT takes place at the interface between the bilayer, especially at the center of the VO$_2$ channel and proceeds from top to bottom. The progression of MIT follows the reverse course.

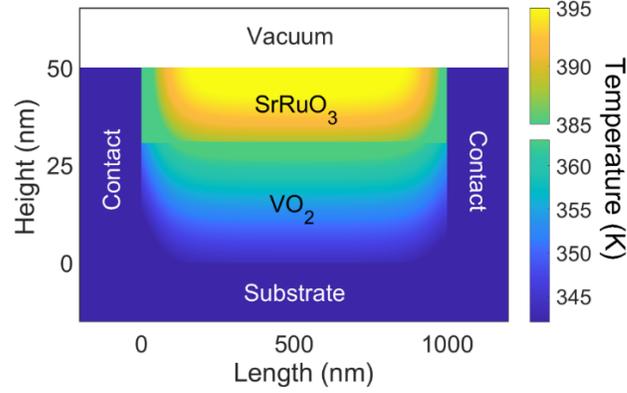

FIG. 6. Temperature profile for the cross section of the channel. The channel dimension was length = 1 μm, width = 1 μm, thickness of $VO_2$ = 30 nm, thickness of $SrRuO_3$ = 20 nm ($V$=1.5 V, t = 10 ns).

## VII. Tunability

Fig. 7 shows the capacitor and current tunability and power spectrum for certain conditions near the highest frequency region in Fig 3(a). Fig. 7(a) shows the frequency tunability at $I_0$=2 mA. About two orders of magnitude difference of frequency can be achieved with the tuning of capacitance from 20 fF to 10 pF. Small changes in capacitance, on the other hand, changes the frequency linearly. Fig. 7(b) shows the frequency tunability at $C$=100 fF. The frequency increased by 3 times from $I_{min}$ to $I_{max}$. The frequency peaks near $I_{max}$ before the current input is too high for the channel to cool down.

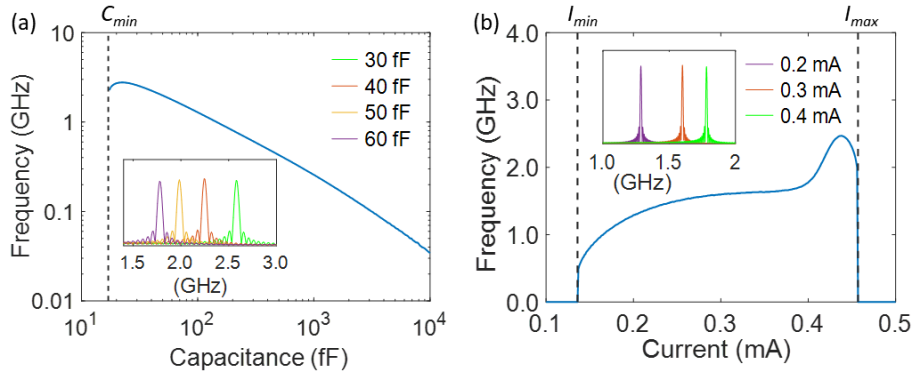

FIG. 7. Capacitor and current tunability and power spectrum near the highest frequency region (a) Frequency tuned by capacitance, $I_0$=2 mA; (b) Frequency tuned by current, $C$=100 fF.